**Reply to "Comment on 'Calculation of two-center nuclear attraction integrals over integer and noninteger *n*- Slater-type orbitals in nonlined-up coordinate systems' "**

Telhat Özdoğan[•]

*Department of Physics, Faculty of Education, Ondokuz Mayis University, Amasya-TURKEY*

**Abstract**

The comments of Guseinov on our paper (T. Özdoğan, S. Gümüş and M. Kara, J. Math. Chem., **33** (2003) 181) are critically analyzed. Contrary to his comments, it is proved that the expansion formula for the product of two normalized associated Legendre functions in ellipsoidal coordinates and the expressions for two-center nuclear attraction integrals have been obtained independently, by the use of basic mathematical rules, not by changing the summation indices of expansion relationships contained in his articles. Therefore, our algorithm is original, is not affected from possible instability problems and can be used in large-scale calculations without loss of significant figures. Meanwhile, it should be mentioned that his comment on the transformation of our formulae into his formulae proves the correctness of our algorithm and therefore can be regarded as a nice sound of science.

**Introduction**

In the comment of I.I. Guseinov [1], it is claimed that the formulae we presented in our recent paper [2] are derived from the relationships in his studies [3-5], by changing the summation indices.

The aim of this study is to prove that the formulae presented in Refs.[2] for the expansion formula for the product of two normalized associated Legendre functions in ellipsoidal coordinates and expressions for two-center nuclear attraction integrals over Slater type orbitals (STOs) are obtained independently, not by changing the summation indices of the formulae in Refs.[3-5], as expressed by I.I. Guseinov.

---

[•] E-mail: telhatoz@omu.edu.tr

**Independent proof of the expansion formula for the product of two normalized associated Legendre functions with different centers**

During the calculation of arbitrary multicenter multielectron molecular integrals over STOs, one shall require formula for the product of two normalized associated Legendre functions centered on points *a* and *b*,

$$T^{l\lambda,l'\lambda}(\theta_a,\theta_b) = P_{l\lambda}(\cos\theta_a)P_{l'\lambda}(\cos\theta_b). \tag{1}$$

The analytical expression for the normalized associated Legendre functions was defined by J. Yasui and A. Saika [6] in terms of factorials. By the use of well-known properties of binomial coefficients, we expressed the following relation for normalized associated Legendre functions as [7]

$$P_{lm}(\cos\theta) = (-1)^{(|m|-m)/2} \sum_k C_{lm}^k (\sin\theta)^{2k+m} (\cos\theta)^{l-(2k+m)}, \tag{2}$$

where $\frac{1}{2}(|m|-m) \leq k \leq E\left(\frac{l-m}{2}\right)$ and

$$E\left(\frac{n}{2}\right) = \frac{n}{2} - \frac{1}{4}\left(1-(-1)^n\right), \tag{3}$$

and

$$C_{lm}^k = \frac{(-1)^k}{2^{2k+m}}\left[\frac{2l+1}{2} F_{l-k}(l+m)F_{k+m}(l-k)F_{2k}(l-m)F_k(2k)\right]^{1/2}. \tag{4}$$

For obtaining the relation for the product of two normalized associated Legendre functions, we need the relations between spherical polar coordinates and ellipsoidal coordinates centered at atoms *A* and *B* separated by a distance *R*, as given below:

$$r_a = \frac{R}{2}(\mu+\nu), \quad \cos\theta_a = \frac{1+\mu\nu}{\mu+\nu}, \quad \sin\theta_a = \frac{[(\mu^2-1)(1-\nu^2)]^{1/2}}{\mu+\nu}, \tag{5.a}$$

$$r_b = \frac{R}{2}(\mu-\nu), \quad \cos\theta_b = \frac{1-\mu\nu}{\mu-\nu}, \quad \sin\theta_b = \frac{[(\mu^2-1)(1-\nu^2)]^{1/2}}{\mu-\nu}, \tag{5.b}$$

Substituting Eqs.(2)-(5) in Eq.(1), we have the following relation for the product of two normalized associated Legendre functions in ellipsoidal coordinates

$$T^{l\lambda,l'\lambda}(\mu,\nu) = P_{l\lambda}\left(\frac{1+\mu\nu}{\mu+\nu}\right) P_{l'\lambda}\left(\frac{1-\mu\nu}{\mu-\nu}\right)$$

$$= \sum_{k} C_{l\lambda}^{k} \left(\frac{[(\mu^2-1)(1-\nu^2)]^{1/2}}{\mu+\nu}\right)^{2k+\lambda} \left(\frac{1+\mu\nu}{\mu+\nu}\right)^{l-(2k+\lambda)}$$

$$\times \sum_{k'} C_{l'\lambda}^{k'} \left(\frac{[(\mu^2-1)(1-\nu^2)]^{1/2}}{\mu-\nu}\right)^{2k'+\lambda} \left(\frac{1-\mu\nu}{\mu-\nu}\right)^{l'-(2k'+\lambda)}$$

$$= \sum_{k\,k'} C_{l\lambda}^{k} C_{l'\lambda}^{k'} \frac{[(\mu^2-1)(1-\nu^2)]^{k+k'+\lambda}}{(\mu+\nu)^{l}(\mu-\nu)^{l'}} (1+\mu\nu)^{l-(2k+\lambda)} (1-\mu\nu)^{l'-(2k'+\lambda)}. \qquad (6)$$

Using an identity

$$(\mu^2-1)(1-\nu^2) = (\mu+\nu)^2 - (1+\mu\nu)^2 \qquad (7)$$

and well-known binomial expansion

$$(\mu+\nu)^{N} = \sum_{m=0}^{N} F_{m}(N)\mu^{N-m}\nu^{m}, \qquad (8)$$

we obtain

$$[(\mu^2-1)(1-\nu^2)]^{k+k'+\lambda} = [(\mu+\nu)^2 - (1+\mu\nu)^2]^{k+k'+\lambda}$$

$$= \sum_{u=0}^{k+k'+\lambda} (-1)^{u} F_{u}(k+k'+\lambda)(\mu+\nu)^{2(k+k'+\lambda)-2u} (1+\mu\nu)^{2u}. \qquad (9)$$

Substituting Eq.(9) in Eq.(6) we find

$$T^{l\lambda,l'\lambda}(\mu,\nu) = \sum_{k\,k'} C_{l\lambda}^{k} C_{l'\lambda}^{k'} \sum_{u=0}^{k+k'+\lambda} (-1)^{u} F_{u}(k+k'+\lambda) \frac{(1+\mu\nu)^{l-(2k+\lambda)+2u}(1-\mu\nu)^{l'-(2k'+\lambda)}}{(\mu+\nu)^{l-2(k+k'+\lambda)+2u}(\mu-\nu)^{l'}}. \qquad (10)$$

Next, using the well-known relation [8]

$$(\mu+\nu)^{N}(\mu-\nu)^{N'} = \sum_{m=0}^{N+N'} F_{m}(N,N')\mu^{N+N'-m}\nu^{m} \qquad (11)$$

in Eq.(10), we obtain

$$(1+\mu\nu)^{l-(2k+\lambda)+2u}(1-\mu\nu)^{l'-(2k'+\lambda)} = \sum_{s=0}^{l+l'-2(k+k'+\lambda)+2u} F_s(l-2k-\lambda+2u, l'-2k'-\lambda)(\mu\nu)^s. \qquad (12)$$

By the use of Eq.(12) in Eq.(10), the expansion formula for the product of two normalized associated Legendre funcitons is obtained as

$$T^{l\lambda,l'\lambda}(\mu,\nu) = \sum_{k,k'}\sum_{u,s} a_{us}^{kk'}(l\lambda,l'\lambda) \frac{(\mu\nu)^s}{(\mu+\nu)^{l-2(k+k'+\lambda)+2u}(\mu-\nu)^{l'}} \qquad (13)$$

where the expansion coefficients are

$$a_{us}^{kk'}(l\lambda,l'\lambda) = C_{l\lambda}^{k} C_{l'\lambda}^{k'} (-1)^u F_u(k+k'+\lambda) F_s(l-2k-\lambda+2u, l'-2k'-\lambda), \qquad (14)$$

and the ranges of the summation indices $k, k', u$ and $s$ are as follows:

$$0 \le k \le E\left(\tfrac{l-\lambda}{2}\right), \qquad 0 \le k' \le E\left(\tfrac{l'-\lambda}{2}\right),$$
$$0 \le u \le (k+k'+\lambda), \qquad 0 \le s \le (l+l') - 2(k+k'+\lambda) + 2u. \qquad (15)$$

In Eq.(11), (12) and (14) $F_m(N, N')$ is defined in Ref.[7].

Meanwhile, similar expressions for the product of two normalized normalized associated Legendre functions in ellipsoidal coordinates can be found elsewhere (see: Eq.(A.12) of Ref.[8] and Eq.(10) of [3]). Whereas the formulae in these references include factorials, our formula includes binomials which have very useful recurrence relations and therefore enable to calculate multicenter integrals with enough speed and accuracy. For not causing new unnecessary polemics, the advantages of our procedure will not be discussed here.

Using expansion formula for the product of two normalized associated Legendre functions (Eq.(13)) in Eqs.(2) and (3) of our paper [2], it is easy to have analytic relations for two center nuclear attraction integrals. It should be noted that nuclear attraction integrals given by Eq.(1) of the comment of Guseinov [1] is not the aim of our recent paper [2]. Recently, we have presented an efficient formula for this type of nuclear attraction integrals in terms of auxiliary functions [9].

Consequently, we note that any physical or mathematical quantity obtained in two different ways can be transformed into one another. The efficiency, accuracy and the usefulness of the procedure for coding in computer are the most important factors in such a case. In this respect, the comment of I.I. Guseinov is important and valuable to confirm our procedures for the expansion formula for the product of two normalized associated Legendre functions and expressions for two-center overlap and nuclear attraction integrals [2,10], which also can be used in the evaluation of multicenter molecular integrals over STOs.

It is seen from the proofs that the formulae presented in Refs.[2,10] are obtained independently, not by changing the summation indices of the formulae in Refs.[3-5], as expressed by Guseinov. Meanwhile, for not causing misunderstanding of readers and publishers, it should be stressed that similar comment made by Guseinov [11] on our paper [10.a] have been replied in Appendix of Ref.[12].